\documentclass[prl,twocolumn,showpacs,showkeys,preprintnumbers,amsmath,amssymb]{revtex4}

 \usepackage{epsf}

\usepackage{amsmath,amsfonts}

\begin{document}
\title{Branes vs. GUTS: challenges for string inspired phenomenology}
\author{David Berenstein$^{1}$}
\email{dberens@physics.ucsb.edu}

 \affiliation{$^1$Department of
Physics, University of California
at Santa Barbara, CA 93106}

\begin{abstract}
This paper studies  the compatibility of having a grand unification scheme for particle physics,
while at the same time having a perturbative string theory description of such a scheme on a D-brane. This is studied in a model independent approach and finds a negative result. Some additional observations related to model building  on branes are made.
\end{abstract}
\pacs{11.25.Uv, 11.25.Wx, 12.10.Dm}
\keywords{D-branes, Grand Unified Theories}
 \maketitle

\begin{center}{\em Introduction}
\end{center}

The standard model of particle physics is a very successful and well established model to describe the 
particle interactions at accessible energies in current high energy experiments. The model contains more than 30 parameters that can be measured,  but they have no explanation of their
relative numerical strengths within the standard model itself. It has been a big dream of particle physics theory to reduce the number of such free parameters and explain various features of the standard model. Grand unified theory (GUT) schemes reduce the number of free parameters because they
relate the different matter fields in the standard model and unify their interactions at some high energy scale. Simple GUT's predict the Weinberg angle remarkably well \cite{Georgi} and this is why they are so interesting.

One of the most salient problems in high energy physics is the hierarchy problem.  Supersymmetry is a technically natural solution of the hierarchy problem.  In recent years, another solution to the hierarchy problem has been proposed \cite{ADD,RS}, based on the idea that the standard model of particle physics might be confined to a brane in higher dimensions, 
and that gravity's strength in four dimensions is diluted by the large extra dimensions volume.
This second type of solution to the hierarchy problem was inspired in the idea of D-branes in string theory \cite{Pol}.  There has also been a very voluminous effort to produce models of the standard model on a D-brane in various setups , see for example \cite{AIQU,BJL,CSU, VW}, and for a recent review see \cite{BCLS}. The main reason for this interest is that D-brane setups provide an {\em in principle calculable} model of particle physics, where perturbation theory calculations can be carried from first principles and one can hope to address the compatibility of string calculations with precision measurements of the standard model. The reason to ask for perturbative physics is obvious. At the electroweak scale --at least with current observations-- the intermediate energy behavior of particle physics is surprisingly well described by perturbation theory.

Now, it behooves us to ask if it possible to have it all: the world on a D-brane, where string perturbation theory is reliable, 
a grand unification scheme that predicts relations between the coupling constants, a prediction of the new energy scale where the string theory effects are important, etc.  Recent ideas 
connected to the landscape (as in solutions of the cosmological constant with fluxes and branes \cite{KKLT}) have indicated that given a naive estimate of the number of vacua in string theory \cite{AD}, the number of solutions is large and the above question should be answered in the affirmative.

In this paper we will ask this question from a model independent point of view:  can perturbative D-brane setups (this is taken to mean that string perturbation theory allows all required Standard model coupling constants at intermediate scales) and Grand Unified ideas be compatible with our knowledge of the 
standard model? The answer I propose to this question is that this is not possible, (a similar conclusion to observations in heterotic models \cite{DS}). I also will discuss what options remain available and give details on why some of them might not be suitable for particle phenomenology.

\begin{center}{\em Basic building blocks}
\end{center}

We are interested in particle physics theories derived from D-brane constructions as a low energy effective description of the dynamics. The details of the models depend a lot on wether we use oriented string or unoriented string constructions. The oriented string models are much simpler. The limit we take is related to the $\alpha'\to 0$ limit on string theory. Only massless, or nearly massless states are kept in the low energy description. The D-branes for the purposes of this paper are thought of as geometric objects in large extra dimensions, but this is not strictly necessary.  We use this visualization as an aid that allows us to talk about them in a geometric language. The same statements can be made using the boundary state formalism in more abstract settings. Here we write well known rules of thumb within the sting theory 
lore that let us describe these models. Here we include them for completeness in case the reader is not familiar with them.

{\em Oriented string models:}

We consider a collection of D-branes on some  geometry that are extended in the four directions we see. There are two cases to consider: If the D-branes are point-like in the extra dimensions, then the low energy physics is described by the massless spectrum of states on  the D-brane. If the D-branes are extended, the low energy physics has Kaluza-Klein harmonics in the extra directions.
We only choose the lowest lying mode for the low energy effective field theory description. For our purposes these modes have to be light enough to fit in the few TeV scale. This is a model dependent issue.

The massless content of coincident D-branes include gauge fields \cite{Pol}. These are protected by gauge invariance from getting a mass. The Higgs mechanism works by separating the 
branes, sometimes also by turning on a Wilson line in the extra dimensions, or by deforming
intersecting branes into a single smooth brane.
The gauge group on a collection of $N$ coincident D-branes is $U(N)$. If we have many D-brane
types, then the gauge group will be of product form $\prod U(N_i)$, where $N_i$ reflects the number of D-branes in each stack. Different stacks generically intersect. When they intersect
one can have massless fields associated to the intersection. For oriented strings, each end of the string carries a fundamental or anti-fundamental representation of the gauge group associated to where the string ends, depending exclusively on the orientation of the ends.
Thus all fields that are charged with respect to two different brane stacks are in bifundamentals. Strings going from one brane to itself are in the adjoint. There are no massless vectors going between different stacks of branes.The matter content can be chiral \cite{BDL}, and the 
total
number of bifundamentals minus ant-bifundamentals is controlled by an intersection index (A good discussion is found in \cite{Douglas}). These also give rise to mixed anomalies whose numerical value depends on this index).

These configurations are usually visualized by a quiver diagram. This is a graph with directed arrows between the nodes (here we follow \cite{DM}). Each node is associated to one stack of D-branes, and has a multiplicity index (the number of branes of the given type). The arrows 
label the (nearly) massless states between them.
If supersymmetry is broken, we should distinguish fermions from bosons in the quiver diagram.
Nearly massless bosons are allowed, as well as tachyons with a small negative mass. This just reflects classical instabilities of the vacuum.

Finally, we get to a description of the allowed couplings. At string tree level, couplings of these modes are associated to disk diagram worldsheets. Tracing Chan-Paton factors gives the result that all tree level string couplings are of the form of a single trace of a closed loop in the quiver diagram, where the closed loop chases arrows from one node to the next. This is a very important point for this paper. If one integrates out massive closed string modes, one can get multi-trace couplings as well, but they are suppressed by the string coupling constant which is taken to be small. These couplings are also usually associated to irrelevant couplings, so they also are suppressed by their energy dependence. Finally, we should note that the allowed perturbative couplings are constrained by allowing one to have a large $N$ limit of the field theory. This is due to the fact that in general we can expect to change the rank of the gauge group by nucleating brane anti-brane pairs. For disk diagrams, this process does not change the numerical values of amplitudes, but it changes $U(N)$ to $U(N+M)\times U(M)$ where $M$ is the number of such pairs and $U(M)$ is the gauge group of the anti-branes. This can be described in terms of classical string theory \cite{Sen, WittenK}. As a note, with the rules above, the number of 
fields has to scale like $N^2$.

 If we concentrate on the couplings of the branes and ignore the terms that involve the anti-branes, this gives us a constraint on the allowed form of the coupling constants. The coupling constants are constants only from a naive perspective of the open string. 
 They are controlled by the geometry (closed string moduli) and their values can vary substantially as we vary the geometry. These couplings carry factors of the string coupling constant $g_s$, which we take to be small. The open string coupling constant appears in all gauge coupling constants. Havig $g_s$ small guarantees that these gauge coupling constants
 are small at the string scale. In this paper we are interested in infrared dynamics associated to the TeV scale. We can keep all coupling constants small in this regime, and still 
 have strong infrared effects like confinement at ultra-low energies.

Finally, cubic anomaly cancellation is easy to describe in these models. Since we are only allowed to have fundamentals, anti-fundamentals or adjoints, the cubic anomaly for $SU(N)$ tells us that the number of fundamentals and anti-fundamentals have to be equal. Graphically
this gives us an easy rule that numbers of chiral fermion arrows going into a node have to be equal to numbers of chiral fermion arrows going out of the node counted with multiplicity.

Mixed anomalies can be cancelled by the Green-Schwarz mechanism. This involves a Higgsing of some of the $U(1)$ fields at low energies, where a closed string mode supplies the longitudinal polarization of the massive photon. Thus the low energy effective gauge group can be a product of $SU(N_i)$. There is always a diagonal $U(1)$ under which nothing is charged, that remains effectively massless after these constraints to low energy physics are added.

{\em Unoriented string models}

For unoriented strings most of the above rules apply. We are allowed many more choices.
 The allowed gauge groups 
are now $U(N)$, $SO(N)$ and $Sp(N)$ and their products. The theory also has to admit a large $N$ limit. Matter fields now can carry also one fundamental index per end of the string. However, because the strings are unorieted, both ends can be fundamental or anti-fundamental, for $SO(N)$ and $SP(N)$ each end carries a vector index. 
Couplings also need to be made from gauge invariant combinations of closed loops in the quiver, and  are also allowed to be multi-trace if we integrate out heavy string modes. The matter content allows also for symmetric and antisymmetric tensor multiplets (not just adjoints from a brane to itself), as well as their complex conjugates if the gauge group is $U(N)$.  The anomaly cancellation condition here has to be calculated on a case by case basis.

\begin{center} \em
Ruling out simple GUT  brane models
\end{center}

There are many GUT models to analyze. In this paper we will look at the simplest ones.
These are $SO(10)$ with matter in the spinor representation and $SU(5)$ models (the minimal GUT). Most other grand unified models contain one of these two as a submodel.

The first one is very easy to rule out. The spinor representation of $SO(10)$ does not show up
as an allowed matter content of D-brane models (it is not a bifundamental).

For the $SU(5)$ model, we can consider two scenarios. Oriented and unoriented string models.
In either one of these models, if we have fundamental fields, there should be another brane where the string ends. Thus these models are not usually unified at the string scale either and sometimes can resemble a flipped $SU(5)$ model instead. However, we can fit the whole standard model set of interactions in $SU(5)$, or a flipped $SU(5)$ model, and this is what we will analyze.

First, let us look at the oriented models.
The oriented ones can not be chiral with respect to $SU(5)$ due to anomaly cancellation: the number of fundamentals has to be equal to the number of anti-fundamentals. One also is not allowed anything other than an adjoint field matter content. This contradicts our low energy observations of parity violation in the standard model, and we would have trouble fitting the
mechanism of mass generation by Higgsing into the field theory analysis.  Higgsing of the $SU(N)$ models of this type can not
produce models that are chiral either. This  is because the intersection index of  any pair of branes that are a subset from the $SU(N)$ brane stack is zero. Thus, for oriented models chirality forces different parts of the standard model gauge group to be located on different stacks of branes.

Now, let us look at an unoriented $SU(5)$ model (or a flipped $SU(5)$ model). The usual fermions as $\bf 10$ and $ \bf \bar 5$ easily fit into allowed matter representations. Anomaly cancellation relates the number of $\bf \bar 5 $ and $\bf \bar 10$ fields into generations. Also, the usual Higgs field assignments are allowed by the unoriented rules. Indeed, various specific models claim that the spectrum found in them is compatible with the standard model assignments \cite{CSU,CKMNW}.

From this point of view, we seem to be able to fit the standard model of particle physics into a grand unification scheme set by D-brane setups. However, we have to look at the models more carefully. In particular, we need to address also the Yukawa couplings of the GUT model. In the usual GUT case the top quark mass is generated by a $\bf 10 \times  10\times 5  $ Yukawa coupling.

According to our rules for D-brane models, this coupling should be built of single (and possibly multi) trace operators in the quiver diagram and should have a smooth large $N$ limit. However, we notice that this is not the case. The generalization to large $N$ of the above coupling is into a coupling of two 
antisymmetric tensors and a fundamental field. Counting boxes of the corresponding Young tableaux, we have five fundamental boxes total (with no anti-fundamental boxes). Thus the $N$-ality (number of boxes in Young tableaux) of the coupling constant is
$5$. Only couplings whose $N$-ality is zero modulo $N$, are gauge invariant, but only those whose $N$-ality is zero are allowed by traces. To make such a coupling gauge invariant we end up  using the totally antisymmetric tensor of $SU(N)$, and this tensor is the problem and the dimension of the coupling constant is not independent of $N$ and there is no smooth large $N$ limit.  
Thus at weak string coupling constant this coupling is perturbatively zero. The 
same argument forbids certain couplings for flipped $SU(5)$ models: those that would be responsible for the $b$ mass.
This is not a problem for $\bf 10 \times \bar 5 \times\bar 5$ couplings that dominate the lepton masses. 
Such $\bf 10 \times  10\times 5 $ couplings can only be generated by non-perturbative effects in string theory (their vanishing was first noticed in \cite{CPS}, see also\cite{AFK}). 
Since these dominate the generation of mass for the heavy quark fields, we would need a new mechanism for understanding these ``non-perturbative effects" without destroying the perturbative reliability of our model.  Usual instantons contribute zero modes from all generation: to get something generation dependent is much harder.  In essence, these models are no better than constructions based on manifolds of $G_2$ holonomy \cite{WG2}. In all of these we can have an intermediate perturbative window to do perturbative analysis in the low energy effective field theory, but it is impossible to calculate the Yukawa couplings strictly from string perturbation theory. 
Such simple perturtbative D-brane models are ruled out. From this point of view, evidence of simple grand unification within perturbative string models prefers perturbative models based on the heterotic string.

Other more complicated arrangements of mass generation are possible in principle. For example, we could base a Yukawa coupling on a $\bf 10\times 10 \times \bar 10 \times \bar 10$ coupling, if we have scalars in the $\bar 10$ (in SUSY models these are available as the superpartners of the $\bf 10$), or of the form $\bf \bar 10\times \bar 10 \times X$, where $X$ is a composite scalar field \cite{AFK}. The same argument above forbids all of these terms from generating quark masses. The reason is that the gauge invariant mass term for the quarks involves a totally antisymmetric index of $SU(3)$:
we need three color boxes in the coupling to make it gauge invariant, forcing $X$ to be composite of colored scalars, whose vev is zero perturbatively as the color $SU(3)$ is unbroken. This argument can also be used to discard non-unified models with some quarks in the antisymmetric of $SU(3)$. These two arguments together are a very strong constraint on perturbative models.

In practice, if one is willing to allow non-perturbative effects to dominate part of the physics, one usually ends up not being able to trust most calculations from first principles. This is particularly challenging for the brane models we are discussing. 
One should just as well study $F$-theory setups and hope that there are some  effects that would lead to some calculable perturbative physics that one can confront with precision electroweak measurements. A field theory attempt along this line of reasoning can be found in \cite{ANV}.

\begin{center}
\em  Some other problems associated to oriented models
\end{center}

We will now quickly go through  simple alternatives based on oriented non-unified models and we will find some other problems with this approach as well.

The first observation we can make from non-unified models is that at high energies the 
$SU(3)$ and $SU(2)$ become $U(3)$ and $U(2)$. The quark doublets are all in the $(\bf 3, \bar 2)$. The strings are oriented, so the representations under $SU(3)$ and the $U(2)$
need to be the same. Anomaly cancellation then tells us that there have to be at least nine chiral doublets that are not charged under color. Since these are chiral, they can only get their mass from the electroweak symmetry breaking. These doublets should therefore be relatively light and can  contribute to the decay width of the $Z$. One can avoid this if they are sufficiently heavy. In this case they would only modify the width via radiative corrections. The high precision of the standard model measurements disfavors this setup, but does not rule it out. In general, the best way of  
seeing these new particles would be via direct production via an intermediate $Z$ or Higgs $h$ 
with enough center of mass energy to be at treshold for real pair production.  Photoproduction of charged particles is also possible.

If we require also supersymmetry, then the situation worsens. Each of the chiral fermions doublets would have a relatively light scalar doublet superpartner. These in general give rise to unsuppressed flavor changing neutral currents via box diagrams and they would affect precision electroweak data. Also,  in some examples, the masses of some charginos would not be able to come from a superpotential Yukawa coupling. They arise instead from higher dimension operators in the lagrangian \cite{BJL}. This brings down the
scale of new physics and in turn affect neutrino masses in a general setup by making them heavier than observed.  This analysis also covers the recent proposal \cite{VW}, although in that particular proposal, fine tunings are allowed that could in principle bypass the objections
presented here.
The case to discard these models is a much weaker case than for GUTS, but it is necessary to work a lot harder to ensure that none of the problems mentioned above crops up. 

One can also try to analyze non-unified unoriented models.
These are very generic and they don't seem to be particularly predictive in a model independent way. To address issues of these models, we have to work with specific constructions. 

I would like to thank S. Kachru, R. G. Leigh, P. Ouyang and H. Verlinde for discussions and correspondence related to this work. Work supported in part by  DOE, under grant DE-FG01-91ER40618

\end{document}